\documentclass{aa}
\usepackage{times,graphicx}

\begin{document}

   \thesaurus{06         % A&A Section 6: Form. struct. and evolut. of stars
              (13.07.1;  % Gamma rays:bursts,
               03.13.6;  % Methods:statistical
               03.13.4)} % Methods:numerical,
             \title{A BATSE-based search for repeaters in the sample of
               gamma-ray bursts detected by the WATCH experiment.}

%   \thanks{}

%   \subtitle{ }

   \author{J. Gorosabel
          \inst{1}
   \and A. J. Castro-Tirado
          \inst{1}
   \and S. Brandt 
          \inst{2}
   \and N. Lund 
          \inst{3}
          }

   \offprints{J. Gorosabel (jgu@laeff.esa.es)}

   \institute{Laboratorio de Astrof\'{\i}sica Espacial y F\'{\i}sica 
              Fundamental (LAEFF-INTA), P.O. Box 50727,
              E-28080 Madrid, Spain
   \and Los Alamos National Laboratory, MS D436, NM 87545, USA
   \and Danish Space Research Institute, Julianne Maries Vej 30 DK-2100 
   Copenhagen $\O$, Denmark.}

   \date{Received date; accepted date}

  \titlerunning{Search for common WATCH-BATSE GRBs}

  \authorrunning{Gorosabel et al.}

   \maketitle

   \begin{abstract}

     This study is the first known attempt to search for gamma-ray burst
     repeaters combining data from gamma-ray experiments flying on board
     different satellites and making use of information derived from the
     bursts detected simultaneously by all the experiments.  The proposed
     method is suitable to correlate GRB data provided by experiments that
     overlap partially or totally in time.

     As an application of this method we have correlated the positions of
     57 gamma-ray bursts observed by WATCH/GRA\-NAT and WATCH/EURECA with
     1905 bursts detected by BATSE.  Comparing the so-called~~~``added
     correlation''~~~between~~~the WATCH and BATSE bursts with that obtained
     with simulated WATCH catalogues, we conclude that there is no
     indication of recurrent activity of WATCH bursts in the BATSE sample.
     We derive an upper limit of $15.8\%$, with a confidence level of
     $94\%$, for the number of WATCH gamma-ray bursts that could represent
     a population of repeaters in the BATSE sample.

   \end{abstract}

%
%  14.Sep.'90: Demo-Vs.
%________________________________________________________________

\section{Introduction}
Despite the advances carried out so far, the origin of the gamma-ray bursts
(hereafter GRBs) remains unknown. The identification of absorption lines in
the optical spectrum of GRB 970508 strongly supports models arising from
sources at cosmological distances (Metzger et al.  1997), but there is
still a lack of knowledge on the mechanisms originating these enigmatic
phenomena.  One of the most important clues that could clarify the nature
of the GRBs would be the detection of a repeater behaviour.

Initial studies showed an apparent evidence of repetition for the BATSE 1B
catalogue (Quashnock and Lamb 1993), suggesting that it would be possible
to have an excess of pairs of GRBs clustered in both time and space (Wang
and Lingenfelter 1995).  This fact was not confirmed by the work carried
out using the BATSE 2B catalogue (Brainerd et al. 1995), although other
studies provided marginal evidence for both temporal and angular clustering
(Petrosian and Efron 1995). Analyses based on autocorrelations with data
from the BATSE 3B catalogue did not find any evidence of repetition
(Bennett and Rhie 1996) and have imposed several constraints to the number
of repeaters (Tegmark et al.  1996).  Finally, recent studies confirm the
lack of repetition in the 4B catalogue and lead to an upper limit to the
repetition rate of $ 0.04$ burst source$^{-1}$ yr$^{-1}$ (Hakkila et al.
1997).

The BATSE~4B catalogue was obtained by the BATSE experiment on board the
{\it CGRO} satellite and contains 1637 GRBs detected from April 1991 to
August 1996 (Paciesas et al. 1998). The BATSE experiment consists of eight
identical detector modules, placed at the corners of the {\it CGRO}
spacecraft and covering energy channels from $\sim 25$ keV to $\sim$ 2 MeV.
It provides error boxes with a minimum radius of $1.6^{\circ}$ (1$\sigma$
confidence level, Fishman et al. 1994). BATSE is detecting bursts at a rate
of 0.8 bursts per day. The bursts are daily added to the so-called Current
GRB Catalogue, which contains the BATSE~4B catalogue plus all bursts
detected after August 1996. When this study was started, the catalogue
contained 1905 sources; this sample constitutes the basis of the present
work.

\begin{figure*}
 \centering
 \resizebox{!}{!}{\includegraphics[width=\hsize,totalheight=10.7cm]{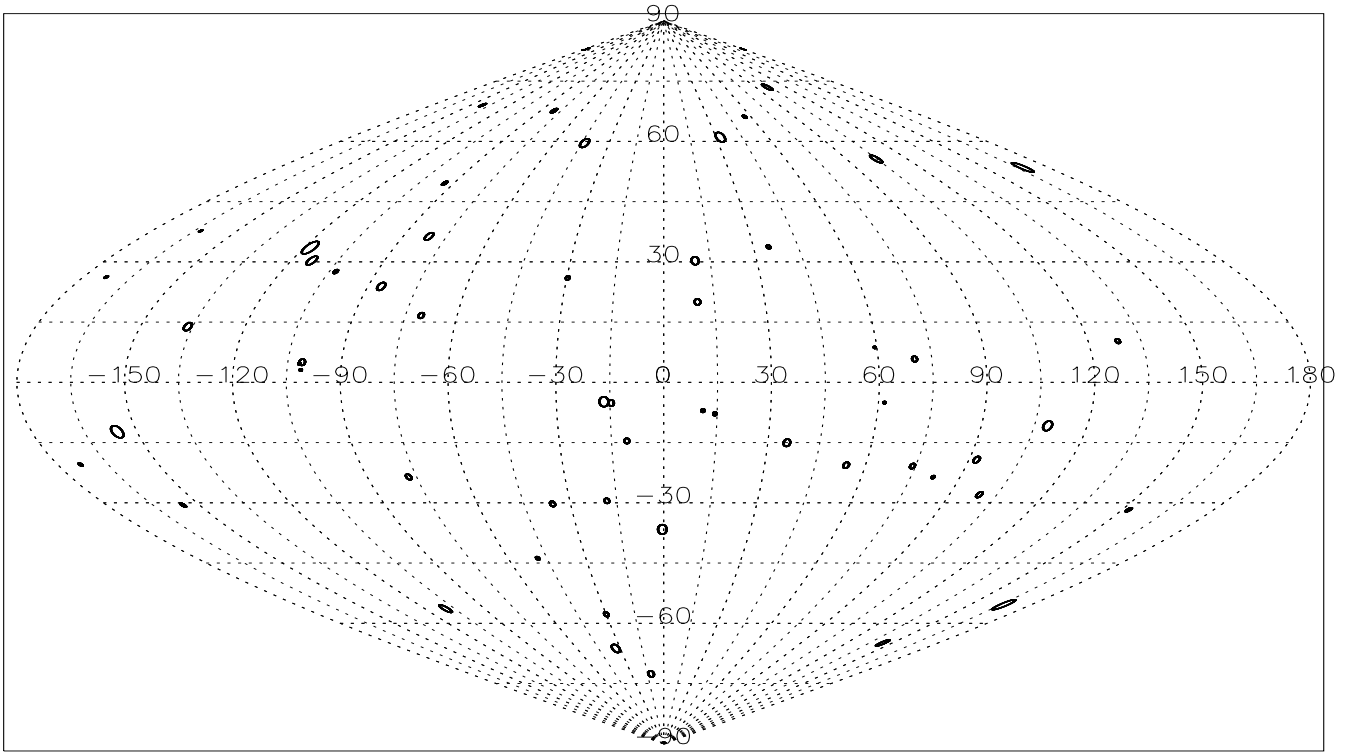}}
\caption{Error boxes for the 57 GRBs detected by WATCH, represented in 
  galactic coordinates. The sample contains 45 GRBs detected by
  WATCH/GRANAT, 10 by WATCH/EURECA and two localized by both experiments at
  the same time. The typical radii of the error boxes are $\sim 1
  ^{\circ}$, with a $3 \sigma$ confidence level.}
\label{figure1}
\end{figure*}

The WATCH X-ray all-sky monitor is based on the rotation modulation
principle (Lund 1986). The instrument has a circular field of view of 4
steradians and an effective area of $\sim$ 30 cm$^2$ (averaged over the
field of view). Position sensitivity is achieved using the rotation
collimator principle, with the collimator grids rotating with a frequency
$\omega$=1 Hz. The phoswich detectors consist of interleaved
scintillator-strips of NaI and CsI crystals.  The geometric area of the
scintillator is 95~cm$^2$.  Four units were mounted on board the Soviet
{\it GRANAT} satellite in a tetrahedral configuration covering the whole
sky, and one unit on board the European Space Agency {\it EURECA}
spacecraft. The total energy range is 8-80~keV, therefore overlapping with
the lower BATSE energy band.  WATCH\-/GRANAT detected bursts in 1990-94 and
WATCH/EURECA in 1992-93, thus both experiments also overlapped in time with
BATSE. One of the main advantages of WATCH was the capability of locating
bursts with relatively small error boxes ($3\sigma$ error radii with $\sim$
1$^{\circ}$) (Brandt et al. 1990).  WATCH\-/GRANAT detected 47 GRBs in this
period and WATCH\-/EURECA 12 (Castro-Tirado et al.  1994, Brandt et al.
1994, Sazonov et al. 1998).  Two GRBs (GRB 920814 and GRB 921022) were
detected by both the WATCH\-/GRANAT and WATCH\-/EURECA experiments.
Therefore, the sample of WATCH GRBs used in this study comprises 57 GRBs:
45 WATCH\-/GRANAT bursts, 10 WATCH\-/EURECA bursts and the above-mentioned
two GRBs.  BATSE also detected 27 of them.  Fig.~\ref{figure1} shows the
sample of 57 WATCH GRBs used in this study.
%<<<<<<<<<<<<<<<<<<< figure 1 >>>>>>>>>>>>>>>>>>>>>>> 

The distribution of time amplitudes for GRBs shows two classes of bursts:
a) durations shorter than $\sim$ 2 s and b) longer than $\sim$ 2 s
(Kouveliotou et al.  1993). It was noticed that the energy spectra of the
short bursts were generally harder than those of the long ones (Kouveliotou
et al. 1993, Lestrade et al.  1993).

The fraction of short events in the WATCH sample is smaller than that in
the 4B catalogue. This fact can be justified by at least three selection
effects:

i) The availability of WATCH for localizing sources is governed by the
rotation speed of the collimator grids (1~Hz). So, a source needs to be
bright enough for at least one rotation of the modulation collimator in
order to be localized, implying a burst duration longer than 1~s. In
contrast, the BATSE experiment is able to detect bursts with durations as
short as 64 ms.

ii) The low energy band of the WATCH experiment ($\sim$8-20~KeV) is
sensitive to the soft GRBs, below the BATSE lower limit ($\sim$25 KeV),
which generally belong to the class of bursts with durations longer than 2
s.

iii) On the other hand, since WATCH is about an order of magnitude less
sensitive than the large-area detectors of BATSE, the WATCH catalogue
contains bursts which are brighter than those in the BATSE sample.

The above three reasons~~explain~~why~~the GRBs~~in the WATCH sample are
longer, softer and brighter than the average BATSE 4B bursts.

This study is the first known attempt to search for repeaters combining
data of $\gamma$-ray experiments flying on board different satellites. The
method proposed makes use of the so-called ``simultaneous bursts'' and is
suitable to correlate GRB data provided by experiments that overlap
partially or totally in time. In the future, this work could also be used
to detect systematic pointing errors between different $\gamma$-ray
experiments, allowing to improve their capability for locating GRBs.

\section{Method}

In this section we outline the methodology proposed to carry out the study.
First, we exclude the simultaneous bursts ($\S$ 2.1) and calculate the
so-called ``added correlation'' function between the WATCH and BATSE
samples ($\S$ 2.2). Afterwards, 1500 WATCH catalogues are simulated ($\S$
2.3) in order to calculate the expected value of the ``added correlation''
function.  Then the distribution of the overlapping function for real and
random overlaps is obtained ($\S$ 2.4) and finally the probability of
having different number of repeaters ($\S$ 2.5) is found.

We consider that there is a common source in both samples when the emission
of a repeater is detected at least twice, once by each experiment and the
detections are separated in time. Thus, the same GRB detected
simultaneously by both experiments \underline{is not} considered as a
common source.  Our study is aimed at searching common sources detected by
both WATCH and BATSE experiments.

\subsection{Simultaneous bursts}
The positions of 27 simultaneous bursts detected by WATCH and BATSE are in
good agreement. If BATSE $1\sigma$ error boxes are considered, there are 20
overlaps with WATCH $3\sigma$ boxes. Instead, if $3\sigma$ error boxes are
taken into account there is only one burst (GRB 920714) that does not
overlap. These 27 bursts were excluded from the BATSE sample of 1905
sources, because they are obviously the same sources detected by WATCH.
Therefore the sample was reduced to 1878 bursts. Nevertheless, the
simultaneous bursts were considered in further calculations ($\S$ 2.4),
because they provide information on the overlapping expected for a repeater
detected by both BATSE and WATCH.

\subsection{The ``added correlation'' estimate}
Recurrence, even in a single case, would be immediately obvious if we had
locations with no errors. However, the locations provided by BATSE and
WATCH, while numerous, have inaccuracies and consequently a statistical
analysis is required to demonstrate, or limit, the presence of common
sources.  If any repeater is present in both catalogues, an excess in the
overlap between the error boxes of both catalogues would be expected. We
define the overlapping function between the $i$-th WATCH and the $j$-th
BATSE error boxes as the following integral over the galactic coordinates
$l$ and $b$:

$$
%\[ 
 c_{ij}=\!
  \left\{ \begin{array}{ll}
  $$ \small \! A F_{j} \int \int W_i(l,l_i,b,b_i) B_j(l,l_j,b,b_j)\, 
    d\Omega  $$
      & \mbox{\scriptsize  if  $d_{ij} < \sigma_{j} + \sigma_{i}$ } \\
  \!  0 & \mbox{\scriptsize  if  $d_{ij} > \sigma_{j} + \sigma_{i}$ } \\ 
  \end{array} \right.
%\]
$$

where $A$ is a normalization factor computed in such a way that $c_{ij}$
remains between 0 and 1. $F_{j}$ is the BATSE exposure correction for the
BATSE $j$-th burst. $l_{i}, l_{j}$ and $b_{i}, b_{j}$ are the galactic
coordinates of the centre of the $i$-th WATCH and the $j$-th BATSE error
boxes, $d_{ij}$ is the distance between the $i$-th WATCH and the $j$-th
BATSE burst, and $\sigma_{i}, \sigma_{j}$ are the $3\sigma$ radii of the
$i$-th~~WATCH~~and~~the $j$-th BATSE error boxes.  $B_j(l,l_i,b,b_i)$ is a
Gaussian-like normalized probability distribution given by the following
expression:

$$B_j(l,l_j,b,b_j)=\frac {\ln(1-s)} {\pi \sigma_{j}^{2}} \exp(~(d_{j}/
\sigma_{j} )^2 \ln(1-s)~)$$

with $s=0.9973$, and $d_{j}$ the distance between the integration point and
the centre of the $j$-th BATSE burst;

 $$\!d_{j}=\! \arccos( \sin (b_{j}) \sin
(b) + \cos (b_{j}) \cos (b) \cos (l-l_{j}) )$$

$\! W_i(l,l_i,b,b_i)$ is~~analogous to~~$B_j(l,l_j,b,b_j)$~~based on WATCH
coordinates. Although we are aware that the errors of BATSE locations do
not follow a single Gaussian distribution (see Briggs et al. 1998), we
consider that, for our purposes, we can extend the Gaussian approximation
from 1$\sigma$ to 3$\sigma$.  This is a very appropriate and useful
approximation which has been frequently used in the past (Fisher et al.
1987, Bennett and Rhie 1996), providing stringent upper limits on the 3B
catalogue (Tegmark et al. 1996).

On the other hand, the error introduced in $c_{ij}$ by considering only
overlaps between 3$\sigma$ error boxes, instead of assuming unlimited error
boxes, is less than 0.1\%, irrelevant for our final conclusions. In the
approximation that $\sigma_{i} << 60^{\circ}$ and $\sigma_{j} <<
60^{\circ}$ (which is quite accurate, since typical values are $\sigma_{i}
\sim 1^{\circ}$ and $\sigma_{j} \sim$ a few degrees), $c_{ij}$
approximately depends on $d_{ij}$ like $\sim \exp { \frac{\ln(1-s)~
    d_{ij}^{2}}{\sigma_{i}^{2}~+~\sigma_{j}^{2}}} $, so it decreases
rapidly when both probability distributions are not close to each other.
$c_{ij}$ provides a measurement of whether both GRBs originated from the
same source or not. Based on the former arguments, we define the ``added
correlation'' $C$ as follows:

$$ C\equiv\sum_{j=1}^{57} \sum_{i=1}^{1878}c_{ij} $$

$C$ is a parameter which is very sensitive to the presence of common
sources in both catalogues. The larger the number of common sources, the
higher the value of $C$ obtained. Our study is based on the comparison of
the ``added correlation'' $C$ calculated for the real WATCH catalogue
(renamed as $C_{BW}$) with those obtained for 1500 WATCH simulated
catalogues (renamed as $C_{j}, j=\{1,2,...,1500\}$).  $C$ is the
generalization for two probability distributions (WATCH and BATSE) of the
$R$ statistics introduced by Tegmark et al. (1996). $C_{j}$ is corrected by
the BATSE and WATCH exposure maps, the first one is taken into account in
the term $F_{j}$ included in the definition of $c_{ij}$, whereas the second
one is considered to simulate the WATCH catalogues for which $C_{j}$ are
calculated.

\subsection{Simulation of WATCH catalogues}
Monte Carlo simulations of 1500 WATCH-like catalogues have been performed.
They provided 1500 values~~for~~$C$~~called $C_{j}, j=\{1,2,...,1500\}$. In
order to determine reliable values for them, the exposure maps of the
WATCH/GRANAT and WATCH/EU\-RE\-CA instruments were taken into account. The
failure of unit number 2 on board {\it GRANAT}, and the limited field of
view and the Earth blockage of WATCH/EURECA, made it that none of the
experiments covered uniformly the sky. The WATCH/GRA\-NAT map shows larger
exposures towards the Galactic centre where\-as the WATCH/EURECA one is
under exposured towards the equatorial poles (Brandt 1994, Castro-Tirado
1994).  If we assume that GRBs occur randomly both in space and time, the
probability of detecting a GRB in a given direction is proportional to the
exposure time spent on that region. Therefore for each simulated set of 57
bursts, 45 of them follow the WATCH/GRA\-NAT exposure map, 10 the
WATCH/EURECA exposure map and the remaining two bursts (representing GRB
920814 and GRB 921022) follow both exposure maps simultaneously.  The
simulated WATCH-like sets have the same error radii than the real WATCH
catalogue.

\subsection{Random and real overlaps}

We call random overlaps to overlaps between the BATSE bursts and the
simulated WATCH events. The random overlaps provide a set of $c_{ij}$ that
follows a distribution so-called $c_{\rm random}$. In order to estimate
such distribution, the value of the overlapping functions $c_{ij}$ are
calculated for all the overlaps between the BATSE sample and 50 simulated
WATCH catalogues. It shows a mean value $ \ < c_{\rm random} > \ =0.0098$
and a deviation $\sigma_{\rm random}=0.035$. $c_{\rm random}$ provides the
expected value of the overlapping function when there is a casual overlap
between two boxes (not due to arise from the same source).  The majority of
the random overlaps shows very low values of the overlapping function
because they tend to occur at the border of the error boxes in the tail of
the probability distribution.

On the other hand, the overlapping function, $c_{ij}$, for each of the 27
BATSE-WATCH simultaneous pairs is calculated. The distribution of these 27
values of $c_{ij}$ is called $c_{\rm real}$. The mean value of the real
overlaps, $ < c_{\rm real} > = 0.28932$, and the deviation $\sigma_{\rm
  real} = 0.22984$. As expected $ <c_{\rm real} > $ is greater than $ <
c_{\rm random} > $.  This fact can be explained taken into account that the
probability distributions due to a single GRB detected by both experiments
tend to be close to each other, compared with two GRBs randomly located in
the same zone in the sky.  Thus, the random overlaps tend to occur in the
tail of the probability distribution, thus forcing $c_{ij}$ to be very low.
Moreover, the lower sensitivity of WATCH in comparison to BATSE implies
that the 27 simultaneous bursts are brighter than the average BATSE bursts,
(as the radii of the error boxes depend on the intensity) and therefore
they have smaller error boxes, thus making $ <c_{\rm real} > $ larger than
$ < c_{\rm random} > $.

The next step is to consider $c_{\rm real}$ as the expected distribution of
$c_{ij}$ for repeaters. This consideration is based on the two following
assumptions:

\begin{enumerate}

\item There is little variation with time on the sensitivity of both
  experiments. A change in the sensitivity imply into differences in the
  sizes of error boxes and thus in the $c_{ij}$ values.  If more accuracy
  is desirable, then it is necessary to know how the sensitivity of both
  instruments evolves, in order to correct the sizes of the error boxes
  depending on the date of detection.

\item The intensities of different bursts from a repeater source do not
  change significantly in time. Therefore the sizes of the repeater error
  boxes remain approximately the same. A more complicated study would deal
  with the time evolution of the repeater sources.

\end{enumerate}

\subsection{Quantification of the number of repeaters}

\begin{figure}[h]
  \resizebox{\hsize}{!}{\includegraphics[angle=-90]{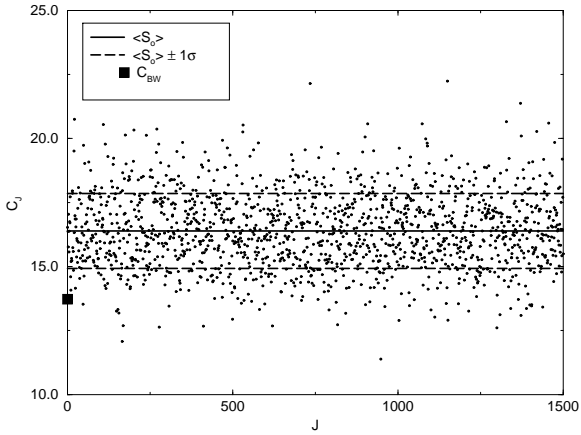}}
\caption{The values of the ``added correlation'' $C_{j}$, for the simulated
  catalogues $j=\{1,2,...1500\}$. The solid line represents $< S_{o} >$,
  the mean value of the ``added correlation'' for the simulated catalogues,
  the long-dashed lines are the $\pm 1 \sigma$ limits. As it is clearly
  seen the real value of the ``added correlation'' $C_{BW}$ (represented by
  the square) is below the $1 \sigma$ limit. Our results are not compatible
  with the presence of common sources, as expected from the graph.}
\label{figure2}
\end{figure}

The set of 1500 $C_{j}$'s calculated using mock WATCH catalogues follows a
Gaussian probability distribution (hereafter called $S_{0}$, see
Fig.~\ref{figure2}). The simulated WATCH catalogues were generated only
using the exposure maps and they only contain accidental overlaps, because
the simultaneous GRBs were excluded from the sample. Therefore $S_{0}$
gives us the expected value of the ``added correlation'' when WATCH and
BATSE catalogues do not share any source.  Assuming that $c_{\rm real}$
represents the expected value of the overlapping function for repeater
sources, we can introduce trial repeaters and construct the $C_{j}$'s
probability distributions $S_{N}$ for different number of repeaters, $N$,
by the following symbolic expression :

$$S_{N} = S_{0} + \sum_{i=1}^N (c_{\rm real}-c_{\rm random})$$

\begin{figure}[h]
\resizebox{\hsize}{!}{\includegraphics[angle=-90]{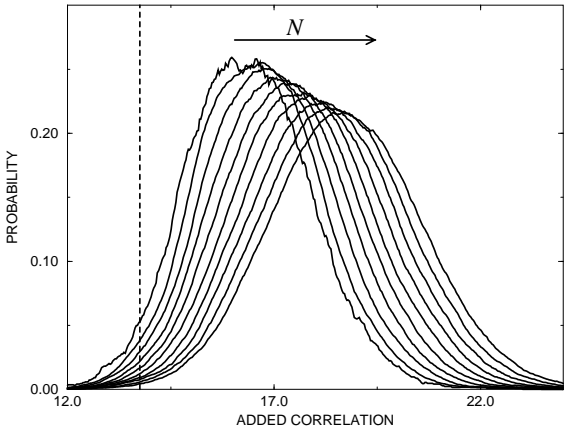}}
\caption{The Gaussian-like curves represent the probability distributions
  $S_{N}, N=\{0,1,2...9\}$, and the vertical dashed line shows the value of
  the BATSE-WATCH ``added correlation''. The intersection of $C_{BW}$ with
  the probability distributions $S_{N}$ provides the set $P_{N},
  N=\{0,1,2...9\}$.}
\label{figure3}
\end{figure}

$N$ being the number of repeaters.  If we take any ``added correlation''
$C_{j}$ of $N=0$ repeaters with a probability given by $S_{0}$, and then we
add the contribution to $C_{j}$ of any repeater with overlapping function
given by $c_{\rm real}$ and subtract the contribution of any random overlap
given by $c_{\rm random}$, we get a new ``added correlation'' $C_{j}$. This
process can be repeated by introducing other real and random overlaps,
providing a new set of $C_{j}$'s.  Once a trial repeater has been
introduced, this set of $C_{j}$'s will follow a different probability
distribution from $S_{0}$, called $S_{1}$.  Thus, $S_{1}$ provides the
expected values of $C_{j}$ when BATSE and WATCH share one source.
Similarly, this method can be applied for $N = \{2,3,...9\} $ repeaters, in
order to obtain the distributions of the ``added correlations'' for
different number of repeaters, $S_{N}, N=\{2,3,...9\}$.  Fig.~\ref{figure3}
shows the probability distributions $S_{N}, N=\{0,1,2,3,...9\}$ obtained
using this procedure.

The intersection between $C_{BW}$ and the distributions $S_{N}$ provides
$9$ values called $P_{N}, N=\{0,1,2...9\}$.  Based on the $9$ values of
$P_{N}$ we can obtain the distribution of $P_{N}$ for $N >9$. Taking into
account that the maximum number of allowed coincidences is $57 \times
1878$, the distribution of $P_{N}$'s can be normalized by imposing $
\sum_{N=1}^{57 \times 1878} P_{N} =1$.  Then, $P_{N}$ provides the
probability that the BATSE and WATCH catalogues share $N$ sources (see
Fig.~\ref{figure4}).

%<<<<<<<<<<<<<<<<<<< figure 2 >>>>>>>>>>>>>>>>>>>>>>> 

%<<<<<<<<<<<<<<<<<<< figure 3 >>>>>>>>>>>>>>>>>>>>>>> 
\section{Results and discussion}

As it can be seen in Fig. \ref{figure2}, the mean value of the ``added
correlation'' for the simulated catalogues, $< S_{o} > =16.40 \pm 1.47$, is
even larger than the ``added correlation'' for the real WATCH and BATSE
catalogues, namely $C_{BW} = 13.72$. This implies that our results agree
qualitatively with the absence of common sources.

The fact that we have preferred to simulate WATCH catalogues instead of
BATSE ones is only due to the computing time, because it is more efficient
to simulate sets of 57 bursts in comparison with groups of 1906 members. In
spite of this fact, the roles of both catalogues were exchanged in order
validate the method, applying the process explained in section $\S$ 2 to 50
BATSE simulated catalogues. Only with 50 catalogues the values obtained for
$C_{BW}$ and $< S_{o} >$ differ by less than 5\% from those obtained when
WATCH catalogues were simulated.

%<<<<<<<<<<<<<<<<<<< figure 4 >>>>>>>>>>>>>>>>>>>>>>> 
\begin{figure}
\resizebox{\hsize}{!}{\includegraphics[angle=-90]{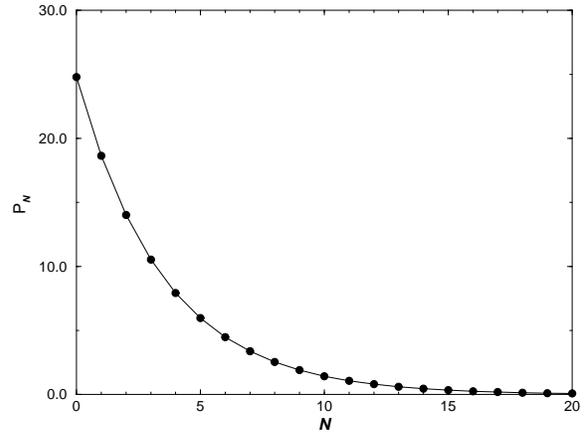}}
\caption{The values of $P_{N}$ for different number 
  of repeaters. The probability of having $N$ repeaters reaches its maximum
  at $N=0$ (no common sources) and decreases rapidly with the number of
  repeaters.}
\label{figure4}
\end{figure}

\begin{table}[ht]
\begin{center}
  \caption{ The first and third column represent 
    the number of repeaters, $N$. The second and the fourth ones give the
    probability of having $N$ repeaters.}
  \begin{tabular}{cccc}
\hline
  \ \   $N$ \ \ & $P_{N}$ (\%)   & \ \  $N$  \ \ & $P_{N}$ (\%) \\
   \hline
   0  &  24.8 $\pm 4.0$    & 5 & 6.0 $\pm 2.0$  \\
   1  &  18.7 $\pm 3.0$    & 6 & 4.5 $\pm 1.7$ \\
   2  &  14.0 $\pm 2.7$    & 7 & 3.4 $\pm 1.5$  \\
   3  &  10.5 $\pm 2.5$    & 8 & 2.5 $\pm 1.1$ \\
   4  &  7.9  $\pm 2.2$    & 9 & 1.9 $\pm 0.9 $ \\
\hline
 \end{tabular} 
\end{center}
\end{table}

The probability distributions $S_{N}$ are shown in Fig.~\ref{figure3} and
the deduced values of $P_{N}$ are given in Table~1 and displayed in
Fig.~\ref{figure4}.  As it is shown in Table~1, $P_{N}$ decreases with $N$,
showing the maximum value when BATSE and WATCH do not share any source.
Thus, our results support the lack of common sources. Furthermore, the
number of common sources is $\leq 9$ with a $94\%$ confidence level (see
Table~1), which means a $15.8\%$ of the whole sample.  This percentage is
similar to the $20\%$ upper limit imposed to the 1B catalogue (Strohmayer
et al.  1994).  The results are also in good agreement with the studies
carried out with the BATSE 3B (Tegmark et al.  1996) and 4B catalogues
(Hakkila et al.  1998), which did not find evidence of repetition.  A
possible reason to explain our results could be due to the different
sensitivity of the experiments, as WATCH is sampling the strongest bursts
and BATSE is also detecting a fainter population. The different populations
of objects found inside WATCH and BATSE error boxes could support this idea
(Gorosabel and Castro-Tirado 1998a, 1998b).

\section{Conclusion}
In this study we have developed a method that allows us to search for GRBs
common to two catalogues of sources, each one based on a different
instrument.  The method makes use of the GRBs detected simultaneously by
both experiments, so it is necessary that the experiments overlap in time.
We have applied~~the~~method~~to the WATCH (WATCH/GRANAT + WATCH\-/EU\-RE\-CA)
and BATSE (BATSE 4B + bursts detected after August 1996) catalogues.
 
We conclude that there is no evidence of recurrent activity of WATCH bursts
in the BATSE sample. We claim (with a $94\%$ confidence level) that no more
than a $15.8\%$ of the 57 GRBs detected by WATCH are present in the sample
of 1905 BATSE bursts (excluding the simultaneous bursts).  However, the
possibility of finding repeaters in each single catalogue cannot be ruled
out. Our results support models which do not predict repetitions of GRBs
(for instance the merging of neutron stars at cosmological distances).

\section{Acknowledgments}
J. Gorosabel wishes to thank B. Montesinos for revising the paper and for
fruitful comments. This work has been partially supported by Spanish CICYT
grant ESP95-0389-C02-02.

\end{document}